\documentstyle[manuscript,osa]{revtex}

\begin{document}
\title{Thermal Fluctuations of Elastic Filaments With Spontaneous Curvature and
Torsion}
\author{S. Panyukov\cite{serg} and Y. Rabin\cite{yit}}
\address{Department of Physics, Bar--Ilan University, Ramat--Gan 52900, Israel }
\maketitle

\begin{abstract}
We study the effects of thermal fluctuations on thin elastic filaments with
spontaneous curvature and torsion. We derive analytical expressions for the
orientational correlation functions and for the persistence length of
helices, and find that this length varies non--monotonically with the
strength of thermal fluctuations. In the weak fluctuation regime, the
persistence length of a spontaneously twisted helix has three resonance
peaks as function of the twist rate. In the limit of strong fluctuations,
all memory of the helical shape is lost.
\end{abstract}

Recent advances in the art of micro--manipulation of molecules led to many
experimental studies of the elasticity of biomolecules such as DNA \cite
{SFB92,PQSC94,SAB96}, chromatin\cite{Bust}, proteins\cite{Gaub}, and rodlike
protein assemblies\cite{MKJ95,Elb,Sack98}. The tacit assumption behind many
of the theories is that the elasticity of these biomolecules is of entropic
origin \cite{MS94,MN98,LGK98} and, consequently, they are modeled as random
walks\cite{PG79}. An alternative approach to the modeling of such systems is
based on the assumption that the origin of elasticity is energy rather than
entropy -- there exists a lowest energy equilibrium configuration with
associated spontaneous curvature \cite{TTH87}, deviations from which give
rise to restoring forces. While such an approach is a straightforward
extension of the usual theory of elasticity of thin rods\cite{Love}, the
description of arbitrary spontaneous curvature and twist involves rather
complicated differential geometry and most DNA--related studies considered
only fluctuations around the straight rod configuration\cite{BM98} (see,
however, references [18] and [19]). Following recent studies on the
elasticity and stability of thin rods with arbitrary spontaneous curvature
and torsion\cite{DR00}, in this work we investigate the effect of thermal
fluctuations on the statistical properties of such filaments. We derive the
differential equations for the orientational correlation functions of the
vectors pointing along the principal axes of the filament, and use them to
calculate the correlators and the effective persistence length of an
untwisted helix. Analytical expressions for the persistence length of a
spontaneously twisted helix are obtained and it is found that this length
varies non--monotonically with the amplitude of fluctuations and exhibits
resonant--like dependence on the rate of twist. We would like to emphasize
that although the present work is motivated by recent studies of
biomolecules, its aim is to construct a theoretical framework for the
description of fluctuating string--like objects that goes beyond current
models of polymer physics, and we do not attempt here to model particular
experiments involving single--molecule manipulation.

A filament of small but finite and, in general, non--circular
cross--section, is modeled as an inextensible but deformable physical line
parametrized by a contour length $s$ ($0\leq s\leq L$ where $L$ is the
length of the filament). To each point $s$ one attaches a triad of unit
vectors ${\bf t}(s)$ whose component ${\bf t}_{3}$ is the tangent vector to
the curve at $s$, and the vectors ${\bf t}_{1}{\bf (}s{\bf )}\ $and ${\bf t}%
_{2}{\bf (}s{\bf )}$ are directed along the axes of symmetry of the
cross--section. Note that ${\bf t}(s)$, together with the inextensibility
condition $d{\bf x/}ds={\bf t}_{3}$, gives a complete description of the
space curve ${\bf x}(s)$, as well as of the twist of the cross--section
about this curve. The rotation of the triad ${\bf t}$ as one moves along the
curve is determined by the generalized Frenet equations 
\begin{equation}
\frac{d{\bf t}_{i}(s)}{ds}=-\sum_{j}\Omega _{ij}(s){\bf t}_{j}(s)  \label{t1}
\end{equation}
Here $\Omega _{ij}(s)=\sum_{k}\varepsilon _{ijk}\omega _{k}(s)$, $%
\varepsilon _{ijk}$ is the antisymmetric tensor and $\left\{ \omega
_{k}\right\} $ are generalized torsions.

The theory of elasticity of thin rods\cite{Love} is based on the notion that
there exists a stress--free reference configuration defined by the set of
spontaneous curvatures and torsions $\left\{ \omega _{0k}\right\} $. The set 
$\left\{ \omega _{0k}\right\} $ together with Eqs. (\ref{t1}) completely
determines the equilibrium shape of the filament. Neglecting
excluded--volume effects and other non--elastic interactions, the elastic
energy associated with some actual configuration $\left\{ \omega
_{k}\right\} $ of the filament is a quadratic form in the deviations $\delta
\omega _{k}(s)=\omega _{k}(s)-\omega _{0k}(s)$%
\begin{equation}
U_{el}\left\{ \delta \omega _{k}\right\} =\frac{kT}{2}\int_{0}^{L}ds%
\sum_{k}a_{k}\delta \omega _{k}^{2}  \label{U1}
\end{equation}
where the bare persistence lengths $a_{1}=E_{1}I_{1}/kT,$ $%
a_{2}=E_{1}I_{2}/kT$ and $a_{3}=E_{2}(I_{1}+I_{2})/kT$ \ ($T$ is the
temperature and$\ k$ is the Boltzmann constant) are expressed in terms of
the elastic moduli $E_{i}$ and the moments of inertia $I_{i}$ about the axes
of symmetry of the cross--section\cite{DR00}. The only limitation on the
validity of Eq. (\ref{U1}) is that deformations are small on microscopic
length scales, of order of the thickness of the filament. The elastic energy 
$U_{el}\left\{ \delta \omega _{k}\right\} $ determines the statistical
weight of the configuration $\left\{ \omega _{k}\right\} $. Calculating the
corresponding Gaussian path integrals we find that $\left\langle \delta
\omega _{i}(s)\right\rangle =0$ and 
\begin{equation}
\left\langle \delta \omega _{i}(s)\delta \omega _{j}(s^{\prime
})\right\rangle =a_{i}^{-1}\delta _{ij}\delta (s-s^{\prime })  \label{cor1}
\end{equation}
We conclude that fluctuations of generalized torsions at two different
points along the filament contour are uncorrelated, and that the amplitude
of fluctuations is inversely proportional to the corresponding bare
persistence length.

The statistical properties of fluctuating filaments are determined by the
orientational correlation functions, $\left\langle {\bf t}_{i}(s){\bf t}%
_{j}(s^{\prime })\right\rangle $. In order to derive a differential equation
for this correlators, we calculate the variation of ${\bf t}_{i}(s)$ under
the substitution $s\rightarrow s+\Delta s$. Integrating Eq. (\ref{t1})
yields in matrix notation (for small $\Delta s$): 
\begin{eqnarray}
{\bf t}(s+\Delta s) &=&\left\{ {\bf 1}-\int_{s}^{s+\Delta s}ds_{1}{\bf %
\Omega }(s_{1})+\frac{1}{2}\int_{s}^{s+\Delta s}ds_{1}\int_{s}^{s+\Delta
s}ds_{2}{\bf \Omega }(s_{1}){\bf \Omega }(s_{2})+\right.  \nonumber \\
&&\left. \frac{1}{2}\int_{s}^{s+\Delta s}ds_{1}\int_{s}^{s_{1}}ds_{2}\left[ 
{\bf \Omega }(s_{1}){\bf \Omega }(s_{2})-{\bf \Omega }(s_{2}){\bf \Omega }%
(s_{1})\right] +...\right\} {\bf t}(s)  \label{tt}
\end{eqnarray}
where the last term appears because of noncommutativity of matrices ${\bf %
\Omega }(s_{1})$ and ${\bf \Omega }(s_{2})$ for different $s_{1}$ and $s_{2}$%
. We multiply the above expression by ${\bf t}_{j}(s^{\prime })$, average
the result and note that for $s+\Delta s>s>s^{\prime }$ the fluctuations $%
\delta \omega _{i}(s_{1})$ and $\delta \omega _{j}(s_{2})$ are uncorrelated
with the fluctuations of ${\bf t}_{i}(s)$ and ${\bf t}_{j}(s^{\prime })$.
This implies that averages of products of ${\bf \Omega }$'s and ${\bf t}$'s
factorize into products of the averages of ${\bf \Omega }$'s and those of $%
{\bf t}$'s. Since the averages of the terms in the square brackets in Eq. (%
\ref{tt}) depend only on $|s_{1}-s_{2}|$ and their difference vanishes, in
the limit $\Delta s\rightarrow 0$ this yields 
\begin{equation}
\frac{\partial }{\partial s}\left\langle {\bf t}_{i}(s){\bf t}_{j}(s^{\prime
})\right\rangle =-\sum_{k}\Lambda _{ik}(s)\left\langle {\bf t}_{k}(s){\bf t}%
_{j}(s^{\prime })\right\rangle  \label{R3}
\end{equation}
where 
\begin{equation}
\Lambda _{ik}=\gamma _{i}\delta _{ik}+\sum_{l}\varepsilon _{ikl}\omega _{0l}%
\text{ \qquad\ with \qquad }\gamma _{i}=\sum_{k}\frac{1}{2a_{k}}-\frac{1}{%
2a_{i}}  \label{gamma}
\end{equation}
The above equations describe the fluctuations of filaments of arbitrary
shape and flexibility and in the following this general formalism is applied
to helical filaments.

Consider a helix without spontaneous twist, such that the generalized
spontaneous curvatures and torsions $\left\{ \omega _{0k}\right\} $ are
independent of position $s$ along the contour. In order to visualize the
stress--free configuration of such a filament, it is convenient to introduce
the conventional Frenet triad of unit vectors which consists of the tangent,
normal and binormal to the space curve spanned by the centerline,
supplemented by a constant twist angle $\alpha _{0}$ which describes the
rotation of the cross--section about this line. The relation between the two
triads is given by $\omega _{01}=\kappa _{0}\cos \alpha _{0},$ $\omega
_{02}=\kappa _{0}\sin \alpha _{0}$ and $\omega _{03}=\tau _{0}$, where $%
\kappa _{0}$ and $\tau _{0}$ are the constant curvature and torsion of the
space curve. The rate of rotation of the centerline about the long axis of
the helix is $\omega _{0}=(\kappa _{0}^{2}+\tau _{0}^{2})^{1/2}$ , the
helical pitch is $2\pi \tau _{0}/\omega _{0}^{2}$ and the radius is $2\pi
\kappa _{0}/\omega _{0}^{2}$. For constant $\left\{ \kappa _{0},\tau
_{0},\alpha _{0}\right\} $, ${\bf \Lambda }$ is a constant matrix and $%
\left\langle {\bf t}_{i}(s){\bf t}_{j}(s^{\prime })\right\rangle $ is given
by $ij-$th element of the matrix $\exp \left[ -{\bf \Lambda }\left(
s-s^{\prime }\right) \right] $. The eigenvalues of the matrix ${\bf \Lambda }
$ can be obtained analytically by solving for the roots of a characteristic
cubic polynomial but the resulting expressions are cumbersome and will be
presented in a longer report. Here we discuss only two limiting cases:

Weak fluctuations, $\sum_{i}\gamma _{i}\ll \omega _{0}$ In this case 
\begin{equation}
\left\langle {\bf t}_{i}(s){\bf t}_{i}(0)\right\rangle =\left( \omega
_{0i}^{2}/\omega _{0}^{2}\right) e^{-s/l}+\left( 1-\omega _{0i}^{2}/\omega
_{0}^{2}\right) \cos (\omega _{0}s)e^{-s/2l-s/2l_{\phi }}  \label{weak}
\end{equation}
with the decay lengths $l$ and $l_{\phi }$ defined by $l^{-1}=\sum_{k}\gamma
_{k}\omega _{0k}^{2}/\omega _{0}^{2}$ and $l_{\phi
}^{-1}=\sum_{k}a_{k}^{-1}\omega _{0k}^{2}/\omega _{0}^{2}$. The physical
meaning of this correlator becomes clear by switching off thermal
fluctuations ($\gamma _{k}=0$). The first term on the rhs of this equation
expresses the fact that the projection of any vector ${\bf t}_{i}$ of the
triad on the symmetry axis of the helix is constant, with magnitude $\omega
_{0i}/\omega _{0}$. The projections on the plane normal to this axis with
magnitudes $(1-\omega _{0i}^{2}/\omega _{0}^{2})^{1/2},$ rotate with angular
rate $\omega _{0}$. In the presence of weak fluctuations, the axis of
symmetry of the helix becomes a random walk and the loss of correlations of
its projections along the axes of the triad is described by the factor $\exp
(-s/l)$. In the second term of Eq. (\ref{weak}), $\exp (-s/2l)$ describes
the loss of correlations of the orientation of the plane normal to the axis
of the helix. The angular persistence length $l_{\phi }$ results from
averaging over the random angle of rotation ($\phi $) with respect to the
axis of the helix, $\left\langle \cos [\omega _{0}s+\phi (s)-\phi
(0)]\right\rangle =\cos \left( \omega _{0}s\right) \exp \left( -s/2l_{\phi
}\right) ,$ with $\left\langle \left[ \phi (s)-\phi (0)\right]
^{2}\right\rangle =s/l_{\phi }$.

Strong fluctuations, $\sum_{i}\gamma _{i}\gg \omega _{0}$. In this limit 
\begin{equation}
\left\langle {\bf t}_{i}(s){\bf t}_{j}(0)\right\rangle =e^{-\gamma
_{i}\left( s-s^{\prime }\right) }\delta _{ij}  \label{titi}
\end{equation}
i.e., the correlators depend only on the bare persistence lengths and memory
about the orientation of the vector ${\bf t}_{i}$ decays over contour
distance $\gamma _{i}^{-1}$. Strong fluctuations destroy all phase coherence
and all correlations between different vectors of the triad and lead to
complete ``melting'' of the helix.

We now proceed to calculate the effective persistence length $l_{p}$ which
controls both the thermal fluctuations of a filament and its elastic
response to external forces. It is defined as the ratio of the rms
end--to--end separation $\left\langle R^{2}\right\rangle $ \ and the contour
length of the filament $L$, in the limit $L\rightarrow \infty $. The
end--to--end vector is defined as ${\bf R=}\int_{0}^{L}{\bf t}_{3}(s)ds$ \
and thus 
\begin{equation}
l_{p}=\lim_{L\rightarrow \infty }\frac{2}{L}\int_{0}^{L}ds\int_{0}^{s}ds^{%
\prime }\left\langle {\bf t}_{3}(s){\bf t}_{3}(s^{\prime })\right\rangle
\label{l}
\end{equation}
A simple calculation yields a result valid for arbitrarily strong
fluctuations:

\begin{equation}
l_{p}=2\frac{\tau _{0}^{2}+\gamma _{1}\gamma _{2}}{\kappa
_{0}^{2}\allowbreak \left( \gamma _{1}\cos ^{2}\alpha _{0}+\gamma _{2}\sin
^{2}\alpha _{0}\right) +\tau _{0}^{2}\gamma _{3}+\gamma _{1}\gamma
_{2}\gamma _{3}}  \label{l1}
\end{equation}
For non--vanishing curvature and torsion, this expression diverges in the
weak fluctuation limit $\gamma _{i}\rightarrow 0$ and the shape of the
filament is nearly unaffected by fluctuations. Non--monotonic behavior is
observed for ``plate--like'' helices, with large radius to pitch ratio, $%
\kappa _{0}/\tau _{0}$. For $\gamma _{i}\rightarrow 0$, the effective
persistence length approaches zero (recall that $L\rightarrow \infty $ in
Eq. (\ref{l})). Thermal fluctuations expand the helix by releasing stored
length and initially increase the persistence length. Eventually, in the
limit of strong fluctuations, the persistence length vanishes again (as $%
\gamma _{3}^{-1})$ because of complete randomization of the filament. The
sensitivity to the constant angle of twist $\alpha _{0}$ increases with
radius to pitch ratio. In the opposite limit of ``rod--like'' helices $%
\kappa _{0}\rightarrow 0,$ the effective persistence length approaches $%
2/\gamma _{3}$ and becomes a function of $a_{1}$ and $a_{2}$ only. Indeed,
since straight inextensible rods do not have stored length, their
end--to--end distance and persistence length are determined by random
bending and torsional fluctuations only, and are independent of twist.

The preceding analysis can be extended to fluctuating filaments with twisted
stress--free states characterized by constant curvature $\kappa _{0},$
torsion $\tau _{0}$ and rate of twist of the cross--section about the
centerline, $d\alpha _{0}/ds$. The relation between generalized torsions $%
\left\{ \omega _{0k}(s)\right\} $ and Frenet parameters $\left\{ \kappa
_{0},\tau _{0},\alpha _{0}(s)\right\} $ is given by $\omega _{01}=\kappa
_{0}\cos \alpha _{0},$ $\omega _{02}=\kappa _{0}\sin \alpha _{0}$ and $%
\omega _{03}=\tau _{0}+d\alpha _{0}/ds$. The calculation of the persistence
length involves the solution of Eq. (\ref{R3}) with periodic coefficients.
Details of the analytical solution will be given elsewhere. For filaments
with circular cross--section $a_{1}=a_{2}$, the persistence length is
independent of twist. In Fig. 1 we present a plot of the persistence length
given in units of the helical pitch $l^{\ast }=l_{p}\omega _{0}^{2}/2\pi
\tau _{0},$ on the dimensionless rate of twist $w^{\ast }=2\omega
_{0}^{-1}d\alpha _{0}/ds$, for a ``plate--like'' helix with large radius to
pitch ratio $\kappa _{0}/\tau _{0}$ and ribbon--like cross--section, $%
a_{1}\ll a_{2}$. Curve 1 corresponds to the case of weak fluctuations, $%
\gamma _{i}\ll \omega _{0}.$ Throughout most of the range, the persistence
length is independent of the rate of twist but a sharp peak appears at $%
d\alpha _{0}/ds=0$ (see insert), accompanied by two smaller peaks at $%
d\alpha _{0}/ds=\pm \omega _{0}/2$. Note that while in the limit of
vanishing pitch, a ribbon--like untwisted helix degenerates into a normal
ring, the cross--section of a twisted helix with $d\alpha _{0}/ds=\pm \omega
_{0}/2$ rotates by $\pm \pi $ and the helix degenerates into a M\H{o}bius
ring. As the amplitude of fluctuations increases, the central peak
transforms into a narrow dip and the two M\H{o}bius peaks become broad
minima (curve 2). Further increase of fluctuations leads to the
disappearance of the M\H{o}bius dips and the central dip becomes broad and
shallow (curve 3). Finally when $\gamma _{i}\gg \omega _{0}$, all dependence
of the persistence length on the spontaneous twist disappears (curve 4).
Note that, as expected from the discussion following Eq. (\ref{l1}), the
persistence length depends non--monotonically on the amplitude of thermal
fluctuations.

In order to understand the origin of the M\H{o}bius resonances we note that
while the effective persistence length is a property of the space curve
given by the Frenet triad, the microscopic Brownian motion of the filament
arises as the result of random forces that act on its cross--section and
therefore are given in the frame associated with the principal axes of the
filament. Since the two frames are related by a rotation of the
cross--section by an angle $\alpha _{0}(s)$, the random force in the Frenet
frame is modulated by linear combination of $\sin \alpha _{0}(s)$ and $\cos
\alpha _{0}(s)$. This gives a deterministic contribution to the persistence
length which, to lowest order in the force, is proportional to the mean
square amplitude of the random force and therefore varies sinosoidally with $%
\pm 2\alpha _{0}(s)$. The observed resonances occur whenever the natural
rate of rotation of the helix $\omega _{0}$ coincides with the rate of
variation of this deterministic contribution of the random force, $\pm
2d\alpha _{0}/ds$.

In summary, we presented a statistical mechanical description of thermally
fluctuating elastic filaments of arbitrary shape and flexibility. We would
like to emphasize that the only limitation on the magnitude of fluctuations
is that they are small on microscopic length scales, and that our model
describes arbitrary deviations of a long filament from its equilibrium
shape. The general formalism was applied to helical filaments with and
without twist. Strong thermal fluctuations lead to melting of the helix,
accompanied by a complete loss of helical correlations. In general, the
persistence length is a non--monotonic function of the amplitude of thermal
fluctuations. Although through most of its range twist has a minor effect on
the persistence length, resonant peaks and dips are observed whenever the
rate of twist approaches zero or equals in absolute magnitude to half the
rate of rotation of the helix.

{\bf Acknowledgment} We would like to thank A. Drozdov for illuminating
discussions. YR acknowledges support by a grant from the Israel Science
Foundation.

\newpage

{\Large {\bf Figure captions}}

\smallskip

Plot of the dimensionless persistence length $l^{\ast }$ as a function of
the dimensionless rate of twist $w^{\ast }$ for a helical filament with
spontaneous curvature $\kappa _{0}=1$, and torsion $\tau _{0}=0.01$ (in
arbitrary units). The different curves correspond to different bare
persistence lengths: (1) $a_{1}=100,$ $a_{2}=a_{3}=5000$, (2) $a_{1}=1,$ $%
a_{2}=a_{3}=100$, (3) $a_{1}=0.1,$ $a_{2}=a_{3}=10$, (4) $a_{1}=0.01,$ $%
a_{2}=a_{3}=10.$ A magnified view of the region of small twist rates is
shown in the insert.

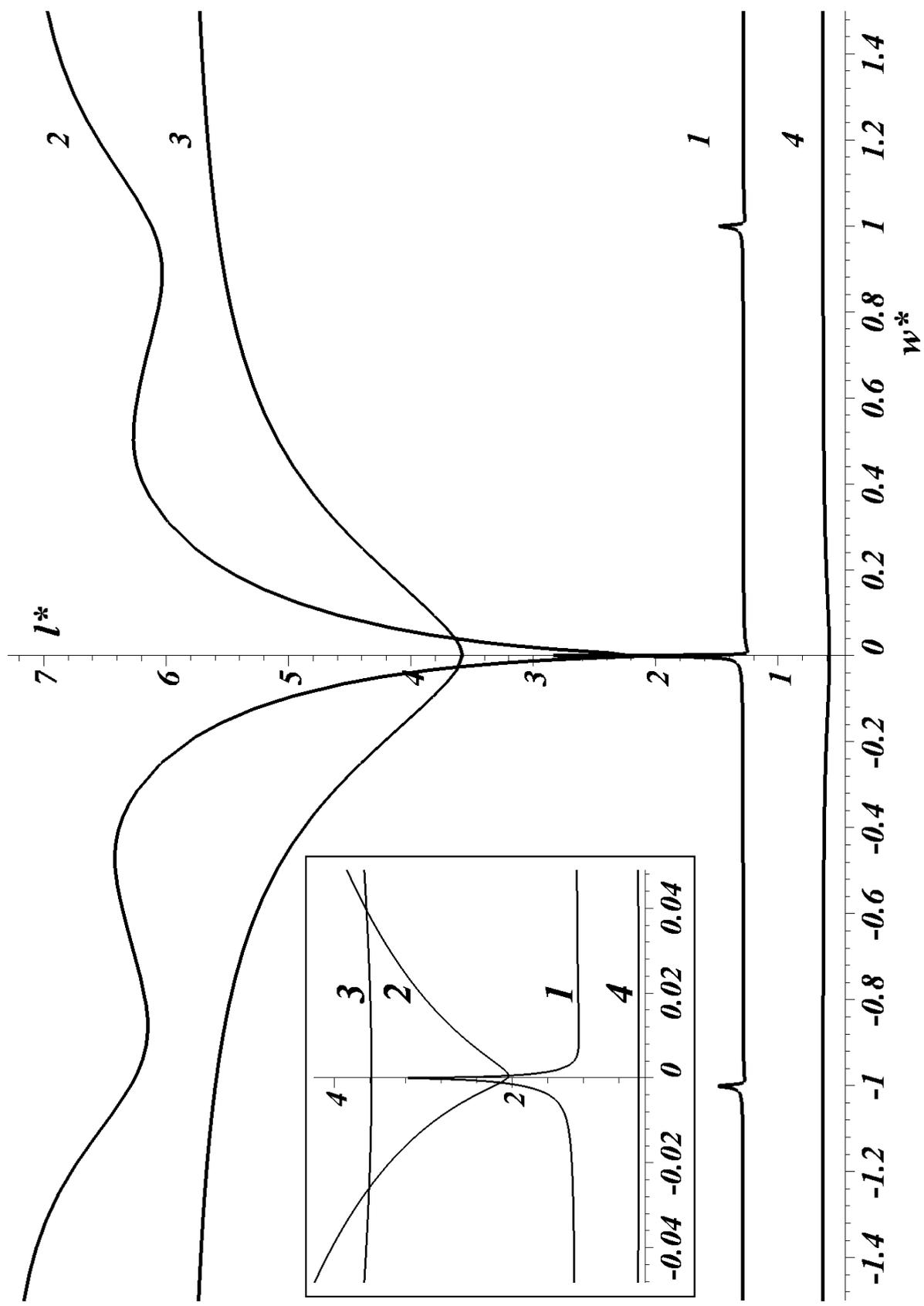
\end{document}